\documentclass{emulateapj}\usepackage{apjfonts}
\usepackage{natbib}
\let\url\relax\usepackage{hyperref}
\newcommand{\doctype}{paper}

\newcommand{\vx}{\ensuremath{\mathbf{x}}}
\newcommand{\vu}{\ensuremath{\mathbf{u}}}
\newcommand{\tc}{\ensuremath{\tau_c}}
\newcommand{\pdf}{\textit{pdf}}
\newcommand{\acvs}{\textit{acvs}}
\newcommand{\p}[1][]{\ensuremath{p_\mathrm{#1}(I)}}

\defcitealias{aime&soummer04}{AS04}

\begin{document}
\journalinfo{ApJ accepted}

\title{Speckle Statistics in Adaptively Corrected Images}
\author{Michael P. Fitzgerald\altaffilmark{1} and James R. Graham}

\affil{Department of Astronomy, 601 Campbell Hall, University of California, Berkeley, CA 94720-3411}
\altaffiltext{1}{fitz@astron.berkeley.edu}

\begin{abstract}
Imaging observations are generally affected by a fluctuating background of speckles, a particular problem when detecting faint stellar companions at small angular separations.  These speckles can be created by both short-lived atmopsheric aberrations and slowly changing distortions in the optical system.  Over the course of a long-exposure image, the combination of many independent realizations of speckle patterns forms a halo in the point spread function (PSF) of characteristic scale $\Delta\theta\sim\lambda/r_0$ (where $r_0$ is the coherence length in the pupil).  While adaptive optics can increase the achievable image contrast, speckle noise remains a major source of random error, which decreases the sensitivity of companion detection observations near the diffraction limit.  Knowing the distribution of the speckle intensities at a given location in the image plane is therefore important for understanding the noise limits of companion detection.

The speckle noise limit in a long-exposure image is characterized by the intensity variance and the speckle lifetime.  In this paper we address the former quantity through the distribution function of speckle intensity.
Previous theoretical work has predicted a form for this distribution function at a single location in the image plane.
We developed a fast readout mode to take short exposures of stellar images corrected by adaptive optics at the ground-based UCO/Lick Observatory, with integration times of 5 ms and a time between successive frames of 14.5 ms ($\lambda=2.2$ $\mu$m).  These observations temporally oversample and spatially Nyquist sample the observed speckle patterns.  We show, for various locations in the image plane, the observed distribution of speckle intensities is consistent with the predicted form.
Additionally, we demonstrate a method by which $I_c$ and $I_s$ can be mapped over the image plane.  As the quantity $I_c$ is proportional to the PSF of the telescope free of random atmospheric aberrations, this method can be used for PSF calibration and reconstruction.
\end{abstract}

\keywords{atmospheric effects --- instrumentation: adaptive optics --- techniques: high angular resolution --- techniques: image processing}

\maketitle

\section{INTRODUCTION}\label{sec:intro}
Wavefronts from astronomical sources are invariably distorted before being imaged by the observer, resulting spatio-temporally fluctuating ``speckles'' of image plane intensity.  Speckle noise limits the dynamic range of long exposures in the vicinity of a bright source~\citep{roddier81,racine99}, hampering observational goals like direct detection of close-in extrasolar planetary systems.

Speckle interferometry~\citep{labeyrie70} and adaptive optics (AO) have succeeded in achieving diffraction-limited spatial resolution in the presence of atmospheric turbulence.  However, contrast levels achievable with these techniques are limited by speckles and are generally orders of magnitude below the dynamic range required for planet detection.  Additional strategies exist to remove speckle noise and increase contrast.  If the speckles are long-lived as in space telescopes, reference PSF subtraction and roll subtraction can be used~\citep{fraquelli04}.   More advanced techniques exist for speckles varying on shorter timescales, such as multiwavelength Simultaneous Differential Imaging~\citep{racine99,marois00,marois04,marois05,lenzen04,biller04}, ``dark'' speckle~\citep{labeyrie95,boccaletti98a}, and Synchronous Interferometric Speckle Subtraction~\citep{guyon04}.

We study the fundamental properties of speckles to illuminate both the limits and design of future instrumentation and observational methodology, in particular high-order adaptive optics coronagraphy~\citep[e.g.][]{sivaramakrishnan01, ford03, trauger03, macintosh04}.

Several authors have developed statistical descriptions of stellar speckles in ground-based images.
In the uncompensated case of the visible and near-IR, speckle interferometry is the main driver of these theoretical and observational studies.  This method uses sequences of short-exposure, spatially well-sampled images to retrieve a single high-resolution image.  The focus of previous statistical studies lies in short-exposure images' spatial power spectra and atmosphere/telescope optical transfer functions which are central to the technique~\citep[e.g.][]{fried66,dainty74,dainty81,roddier81,aime86,vernin91}.

In contrast to techniques which increase resolution of uncompensated images through post-processing, AO images possess a diffraction-limited core formed by spatially coherent light in the pupil.  The superposition of this coherent light with residual wave aberrations modifies the character of the speckle distribution from the uncompensated case.  The description of speckle statistics in both regimes can be related to the study of laser speckles~\citep{goodman75,goodman_statopt}, the theoretical statistical formalisms of which have been previously applied to AO-corrected astronomical speckle patterns~\citep[hereafter AS04]{canales&cagigal99a,soummer&aime04,aime&soummer04}.

For companion detection, we are interested in the distribution of intensity in the image plane.  In high-Strehl-ratio images, diffraction of wave aberrations results in amplification of speckles near Airy maxima, hence the phenomenon of ``speckle pinning''~\citep{bloemhof01,sivaramakrishnan02,bloemhof03,perrin03}.  As noted by~\citetalias{aime&soummer04}, this behavior is described by the theoretical distribution of intensity originally derived by~\citet{goodman75}~\citep[see also][]{goodman_statopt}.
\citet{cagigal&canales01b} observed this distribution in a laboratory setting, using a spatial light modulator to simultaneously simulate turbulence and partial correction with AO.

This \doctype\ presents on-sky results consistent with predictions of the theoretical speckle statistics for compensated astronomical imaging through the atmosphere.  Section~\ref{sec:inten_dist} partially recapitulates the treatment of the intensity distribution given by~\citetalias{aime&soummer04}.  Section~\ref{sec:obs} describes observations of short-exposure stellar images under moderate correction.  Analysis and discussion of these results in light of direct imaging techniques follow in sections~\ref{sec:analysis} and~\ref{sec:conclusions}.

\section{PROBABILISTIC DESCRIPTION OF INTENSITY}\label{sec:inten_dist}
Here we characterize the probability distributions of image plane intensity in the case of an unresolved astronomical source subject to random wavefront aberrations.  In this section, we follow the development of the probability density function presented by~\citetalias{aime&soummer04}, adopting their notation.  The problem has been studied elsewhere to the same conclusions by~\citet{goodman75,goodman_statopt},~\citet{canales&cagigal99a}, and~\citet{soummer&aime04}.

\subsection{Intensity in the Image Plane}\label{subsec:int_implane}
We wish to study the properties of intensity in the image plane by considering an undistorted light wave from an astronomical source, as well as aberrations of the wave introduced by the atmosphere and telescope.  Studying the superposition of the perfect wave and the aberrations allows us to conveniently express intensity statistics useful for companion detection.  Here, we consider the case of a narrow-band wave ($\Delta\lambda/\lambda\ll 1$) which has entered the telescope, passed through any corrective optics (e.g. AO system), and further propagated into the entrance pupil of an imaging system.  We ignore polarization and treat the light as a scalar wave.  We represent the wave at a given location by a complex amplitude, which describes the envelope function of the oscillating (scalar) electric field.

At a given time $t$ and position $\vu$ in the entrance pupil of the imaging system, the complex amplitude of the wave is represented by $\Psi_1(\vu,t)$ and can be decomposed into the superposition of an undisturbed wave of amplitude $A$, which is constant in space and time, and random wave aberrations $a(\vu,t)$,
\begin{equation}\label{eq:pup_wf}
\Psi_1(\vu,t) \equiv \left[A+a(\vu,t)\right]P(\vu).
\end{equation}
The aberrations $a(\vu,t)$ are time-varying fluctuations in the residual wave amplitude arising from the partially-corrected distortions.  The pupil function (unity inside the pupil, zero outside) and static aberrations are encapsulated in $P(\vu)$.  With this definition we may take $A$ as a positive real number without loss of generality.

The wave in the image plane $\Psi_2(\vx)$ is the Fourier transform $\mathcal{F}\{\cdot\}$ of the wave in the pupil plane,
\begin{eqnarray}
\Psi_2(\vx,t)&=& \mathcal{F}\lbrace\Psi_1(\vu,t)\rbrace = A\mathcal{F}\lbrace P(\vu)\rbrace + \mathcal{F}\lbrace a(\vu,t)P(\vu)\rbrace, \label{eq:im_wf} \nonumber\\
 &=& C(\vx)+S(\vx,t).
\end{eqnarray}
The resulting image plane wave is represented as a superposition of two complex wave amplitudes. $C$ represents the amplitude of the temporally static portion of the image plane wave, while $S$ represents the amplitude arising from random aberrations.

The intensity is the squared modulus of $\Psi_2(\vx,t)$,
\begin{equation}\label{eq:im_int}
I(\vx,t) = \vert\Psi_2(\vx,t)\vert^2 = \vert C(\vx)\vert^2 + \vert S(\vx,t)\vert^2 + 2\mathrm{Re}\left[C^*(\vx)S(\vx,t)\right],
\end{equation}
where $^*$ represents the complex conjugate.  Additionally, one can define $I_c$ and $I_s$ as the time-averaged intensities corresponding to the coherent and speckle wavefront amplitudes in the image plane:
\begin{eqnarray}
\langle I(\vx,t)\rangle &=& I_c(\vx)+I_s(\vx) \label{eq:mean_int} \\
I_c(\vx) &\equiv& \vert C(\vx)\vert^2 \\
I_s(\vx) &\equiv& \langle\vert S(\vx,t)\vert^2\rangle \label{eq:i_s}
\end{eqnarray}
The final intensity term on the right of equation~(\ref{eq:im_int}) is an interference term resulting from the superposition of the $C$ and $S$ waves, which has zero mean, and thus does not enter into equation~(\ref{eq:mean_int}).

Although $I_c(\vx)$ and $I_s(\vx)$ are positive definite quantities, we stress that energy is conserved when equation~(\ref{eq:mean_int}) is integrated over the image plane.  Indeed, when adding perturbations to the system, energy conservation constrains the value of $A$.  Consider the case of pure phase perturbations, 
\begin{equation}
\Psi_1(\vu,t)=e^{i\phi(\vu,t)}P(\vu),
\end{equation}
which have no effect on the spatially-integrated mean intensity.  Qualitatively, as a direct result of the definition in equation~(\ref{eq:pup_wf}) and energy conservation,
an increase in the scale of $\phi(\vu,t)$ decreases the value of $A$ and the scale of $\int I_c(\vx)d\vx$.

In this \doctype, probabilistic descriptions are formed of the intensity at a single location in space, $I(t)$, omitting $\vx$ for brevity.  We describe the intensity as a stochastic process -- a sample function $I(t)$ is observed, which is drawn from the overall ensemble of such functions.  For simplicity, we assume that the aberrations driving the intensity fluctuations constitute an ergodic process, so that we may take ensemble averages $\langle\cdot\rangle$ in equations~(\ref{eq:mean_int}) and~(\ref{eq:i_s}) to equivalently represent time averages.  In reality, atmospheric fluctuations are not temporally stationary, and we could redefine $A$ as a function of time, and $I_c$, $I_s$ as the ensemble-averages of temporal functions.

\subsection{The Distribution of Intensity}\label{subsec:int_dist}
The behavior of speckles can be described via the probability distribution of intensity, characterized by \p, the probability density function (\pdf).  This is a first-order description of the stochastic process for $I$ in which time is marginalized.

We assume that in equation~(\ref{eq:im_wf}), the image-plane amplitude $\Psi_2$ is formed by the linear combination of a large number of independent phasors from the pupil plane.  From the central limit theorem, this addition results in Gaussian statistics for the real and imaginary parts of the image plane amplitude, regardless of the particular statistics governing the pupil plane phasors.  The Gaussian \pdf\ of the image plane phasors is used to compute the \pdf\ of the intensity, \p.

The form of \p\ follows from steps involving a change of variables from the phasor components.  Originally derived by~\citet{goodman75}, the resulting function is the modified Rician (MR) distribution,
\begin{equation}\label{eq:modric}
\p[MR] = \frac{1}{I_s}\exp\left(-\frac{I+I_c}{I_s}\right)I_0\left(\frac{2\sqrt{II_c}}{I_s}\right),
\end{equation}
where $I_0(x)$ denotes the zero-order modified Bessel function of the first kind.  For $I_c=0$, the distribution becomes the exponential statistics of `pure' speckle, of primary interest in speckle interferometry.

Intensity fluctuations play a critical role in the signal-to-noise ratio of direct imaging observations.  \citet{goodman75} showed that the mean and variance of $I$ are
\begin{eqnarray}
\mu_I &=& I_c+I_s \label{eq:modric_mn} \\
\sigma_I^2 &=& I_s^2+2I_cI_s. \label{eq:modric_var}
\end{eqnarray}
From equation~(\ref{eq:modric_var}), we see that the variance arising from pure speckle (the first term on the right-hand side of the expression) is augmented by the presence of the coherent wave.

We have expressions describing the statistics of image plane intensity in an idealized absence of photon and detector noise.  The effect of photon noise on equations~(\ref{eq:modric}) and~(\ref{eq:modric_var}) has been studied elsewhere~\citep{canales&cagigal99b,aime&soummer04,soummer&aime04}.  The addition of incoherent light (e.g. from a planet) is addressed by~\citetalias{aime&soummer04}.
Detector read noise can be a significant factor in short-exposure images.  We can model read noise as the sum of a Gaussian random variable with the actual speckle intensity.  We take the read noise distribution to be zero-mean with standard deviation $\sigma_\mathrm{rn}$.  Denoting this the MRG model, the \pdf\ for the resulting observable is broader than the MR case, and is given by the convolution of equation~(\ref{eq:modric}) with a Gaussian \pdf,
\begin{equation}\label{eq:modric+g}
\p[MRG] = \p[MR]\star \p[G].
\end{equation}

\section{OBSERVATIONS}\label{sec:obs}

\subsection{Measurements and Data Processing}

On 2004 July 31, we observed the star Sadalmelik ($\alpha$ Aqr, G2Ib), chosen for its  apparent brightness in the K band ($m_K=0.59$, $2.5\times 10^{7}$ photons m$^{-2}$ $\mu$m$^{-1}$ s$^{-1}$).  We obtained compensated short-exposure images under good weather conditions using the AO system of UCO/Lick Observatory's 3-meter Shane Telescope on Mt. Hamilton~\citep{bauman02}.  There are 61 actuators on the deformable mirror used for wavefront control.  The AO system was run at 500 Hz in a closed feedback loop.

Rapid sequences of short-exposure images were obtained with IRCAL, the infrared camera which images the AO-corrected light onto a 256-by-256 Rockwell Semiconductor PICNIC array~\citep{lloyd00}.  Detector readout is driven by a digital signal processor within an Astronomical Readout Cameras, Inc. electronics system~\citep{leach00}.  The observations were made through a narrow-band Br$\gamma$ filter (2.16 $\mu$m, $\Delta\lambda/\lambda\sim 1\%$), where the detector is Nyquist sampled by 75.6 mas pixels.

We developed an array readout mode optimized for minimal delay between exposures in order to accurately measure the spatial and temporal distribution of stellar speckle patterns.  Typical mode parameters were for correlated-double-sampling (CDS) a 32$\times$32 pixel ($2\farcs 4\times 2\farcs 4$) subarray with 5 ms integration times.  With these settings, exposures were obtained every $\Delta t=14.5$ ms, and in 13 seconds a sequence of 900 images could be obtained before writing to disk.  As demonstrated below, time spent in overhead between exposures (pixel `dead time', about 9.5 ms) is unimportant, as it is shorter than the speckle decorrelation time ($t_\mathrm{dead}<\tc$).  The \emph{rms} CDS read noise $\sigma_\mathrm{rn}$ is measured to be 45 $e^-$ in this mode.  Photon noise fluctuations do not strongly affect the statistics of these data.

The image sequences were processed before analysis, including bias-subtraction and flat-fielding.  Tip/tilt error affects image plane intensity statistics, a fact illustrated by considering the extended Mar\'echal approximation: the Strehl ratio (and by extension, total coherent intensity) decreases exponentially with the phase variance.  We removed the residual tip/tilt error, reducing phase variance, to maximize the range of $I_c$.  After a centroid measurement, each image was shifted and resampled with a damped-sinc interpolator.  The observed 2-D \emph{rms} tip/tilt error was .43 pixels (32 mas).

\begin{figure*}
\plotone{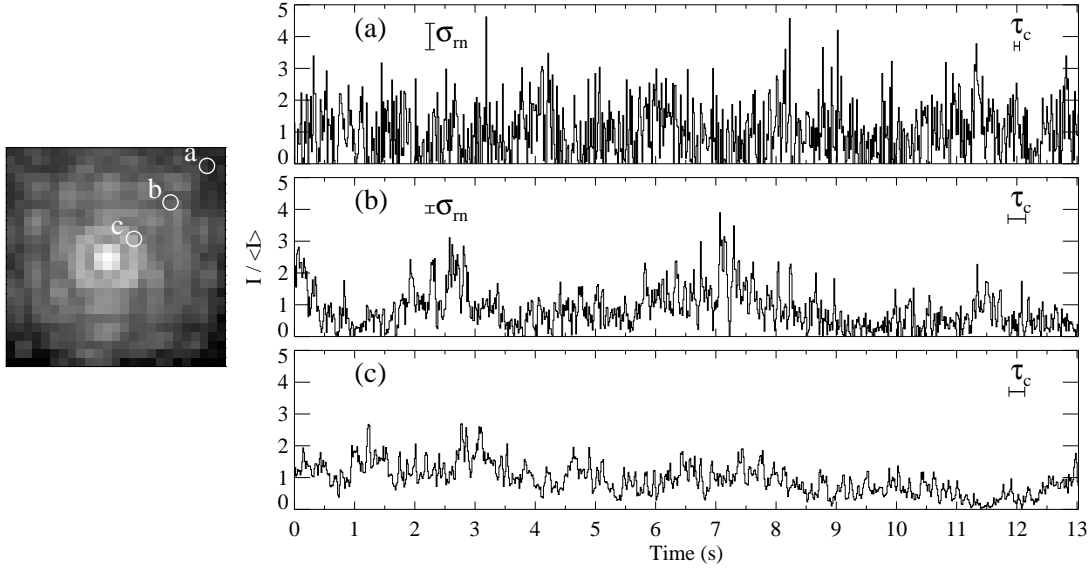}
\caption{The long-exposure image (left, log scale), and a plot of observed intensity as a function of time for three selected pixels (right), the encircled locations ($a$), ($b$), and ($c$).  The locations were chosen to highlight the diversity of speckle intensity distribution as a function of image plane position, moving from the PSF halo to the first Airy ring.  For each pixel, 900 measurements are shown at a sampling period of 14.5 ms, which consists of 5 ms integration and 9.5 ms of dead time.  To illustrate differences in speckle fluctuations with field position, the intensity for each location is normalized by its sample mean, which was 54, 200, and 1800 $e^-$ respectively for ($a$), ($b$), and ($c$).  The scale of fluctuations relative to the mean intensity decreases from locations ($a$) to ($c$).  Read noise and decorrelation time are shown as scale bars.  The read noise at location ($c$) is negligible.}
\label{fig:timeseries}
\end{figure*}

In section~\ref{sec:analysis}, we will examine the distribution function for speckle intensities at each pixel using the measured time series.  Examples of these time series for three locations in the image plane are shown in Figure~\ref{fig:timeseries}.  The locations were chosen to highlight the diversity in intensity distribution  as a function of location in the image plane.  Large deviations from the mean intensity are visible at location ($a$), while the data in location ($c$), close to the PSF core, do not show such large excursions.

\subsection{Temporal Characteristics}\label{subsec:obs_temp}
We wish to analyze our measurements in light of the probability distribution functions for the intensity random variable.  The statistical tests we will employ in~\S\ref{subsec:hyp_test} require that the tested data be drawn from independent random variables.  However, we observe the intensity as a realization of a temporal random process, and correlations exist between intensities at different times.  In order to perform tests concerning the distribution of independent values of intensity, we must determine the timescale over which samples of $I(t)$ can be considered independent.  We assume temporal stationarity for the duration of the observations, so that we may take the statistical properties to be constant in time.

At a single spatial location in the tip/tilt-removed image sequence, the number of independent samples can be characterized by the speckle decorrelation time, $\tc(\vx)$.
Previous definitions of \tc\ have been motivated by the the desire to compute optimal exposure times in speckle interferometry~\citep*[and references therein]{roddier82}.  The decorrelation time is derived from the temporal autocovariance function $\mathcal{C}_I(\tau)$.  It can be the e-folding value of this function, as in~\citet{scaddan&walker78},~\citet{roddier82}, and~\citet{vernin91}.  It can also be defined by the equivalent width,
\begin{equation}\label{eq:tau_c}
\tc = \sigma_I^{-2}\int_{-\infty}^{\infty}\mathcal{C}_I(\tau)d\tau,
\end{equation}
following~\citet{aime86}, who found their empirical $\mathcal{C}_I(\tau)$ was fit by the sum of two Lorentzian functions.
The equivalent-width definition assumes no prior knowledge about the shape of the autocovariance function, and has the advantage of using information contained in the function's tail.  Decorrelation times \tc\ computed from equation~(\ref{eq:tau_c}) are generally longer than those computed by $e$-folding~\citep{aime86}.  This definition also has the property that the variance of the intensity integrated over time $T\gg\tc$ is given by $\sigma_I^2T/\tc$.

In practice, we compute \tc\ for each pixel, approximating the integral as the sum of an estimator for the autocovariance sequence (\acvs) over the interval $\pm 35$ samples ($\pm 0.5$ s).
We restrict the summation domain because the \acvs\ estimate becomes more uncertain with increasing lag $\tau$.
Estimating the error in \tc\ is complicated by the large correlations between terms in the \acvs\ estimate, and it suffices for the analysis in \S\ref{subsec:hyp_test} to neglect its calculation.
Limiting the summation domain in the calculation of \tc\ generally results in an underestimate of the true decorrelation time.  Finally, note that the effect of read noise is not removed from the \acvs\ in the \tc\ calculation.  In this case, \tc\ will be lower than the speckle-only decorrelation timescale in regions of significant detector noise.

\begin{figure}
\plotone{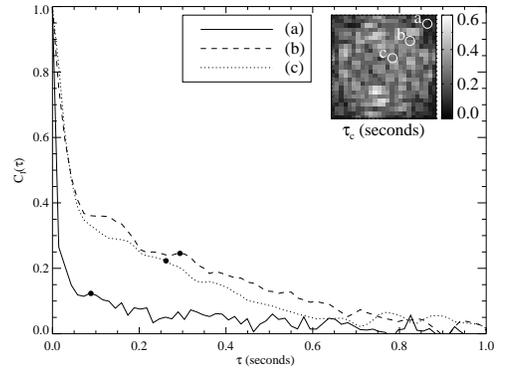}
\caption{Autocovariance sequence (\acvs) estimates for the time series intensity data from pixels ($a$)-($c$) in Figure~\ref{fig:timeseries}. The abscissae of the filled circles denote the decorrelation times \tc\ computed from the \acvs\ estimates.  The \tc\ characterize the number of independent speckle realizations in a given time period, and are observed to vary systematically with field position as illustrated by the inset map.  Sampling effects are minimized with the short sampling period of 14.5 ms.}
\label{fig:acvs}
\end{figure}

Figure~\ref{fig:acvs} shows the \acvs\ estimates for the selected locations from Figure~\ref{fig:timeseries}, along with the \tc\ given by the aforementioned procedure.
As shown in the inset of Figure~\ref{fig:acvs}, the \tc\ are observed to vary with field position $\vx$.  Analysis of the spatial dependence of speckles' temporal decorrelation is deferred for future work.

The measured decorrelation times are generally many times the sampling rate $\Delta t$, which ensures that we may disregard the effect of finite integration time and pixel dead time on the analysis of the observed intensity distributions.
The median decorrelation time was 175 ms, and the first and third quartiles of the measured $\tc(\vx)$ distribution are 95 ms and 270 ms, respectively.

As in~\citet{aime86}, the forms of the \acvs\ estimates in Figure~\ref{fig:acvs} are qualitatively consistent with the sum of two Lorentzian functions.  A unique decorrelation timescale characterizes the width of each component Lorentzian.  For locations ($a$)-($c$), the fast timescale varies between 20 and 80 milliseconds while the slow timescale lies between .8 and 1.2 seconds.

\section{ANALYSIS}\label{sec:analysis}

\subsection{Testing the Distribution}\label{subsec:hyp_test}

The observed distribution of speckle intensities can act as a consistency check against the modified Rician distribution, \p[MR]\ defined in equation~(\ref{eq:modric}).  Neglecting photon counting fluctuations, we expect the observed intensities to be the sum of a modified Rician random variable and an independent Gaussian random variable for the read noise, with distribution \p[MRG]\ given by equation~(\ref{eq:modric+g}).  We adopt as a null hypothesis that the distribution of intensities follows the model distribution, with parameters $\hat{I_c}$ and $\hat{I_s}$ estimated from the data (where the caret denotes an estimate).

\begin{figure}
\plotone{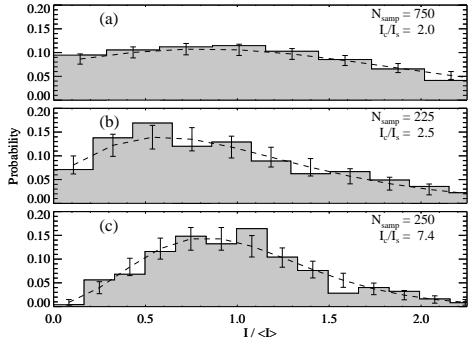}
\caption{Histograms of intensity with best-fit model distribution function (dashed curve) for the data of locations ($a$)-($c$) shown in Figures~\ref{fig:timeseries} and~\ref{fig:acvs}.  Also shown are 1-$\sigma$ errors for the model distribution.  As in Figure~\ref{fig:timeseries}, intensity is normalized by the sample mean.  Inset are the number of samples in this histogram (after resampling the original timeseries by $\Delta t'=\tc$), and the ratio of parameters $\hat{I_c}$ and $\hat{I_s}$ characterizing the model distribution.}
\label{fig:distfits}
\end{figure}

For each pixel location, the observed sequence of images gives a measure of the intensity as a function of time.  Data from five consecutive 900-exposure sequences were combined into the same histogram.  Section~\ref{subsec:obs_temp} describes temporal correlations in the observed sequences, which arise from correlations in the perturbations $a(t,\vu)$.  To mitigate the effect of temporally correlated data in hypothesis tests which assume independent measurements, we formed the histogram after decimating each intensity sequence at intervals of the pixel's decorrelation time, $\Delta t'=\tc$.  The data in between these intervals were discarded, as were values falling below zero.  Examples of the resulting histograms are shown in Figure~\ref{fig:distfits} using the data for locations ($a$)-($c$) from Figures~\ref{fig:timeseries} and~\ref{fig:acvs}.

An iterative least-squares fit of the model distribution to the histogram estimates the parameters $\hat{I_c}$ and $\hat{I_s}$, holding the read noise $\sigma_\mathrm{rn}$ fixed.  We perform this fit at each location in the image.  Best-fit models for distributions of locations ($a$)-($c$) are also shown in Figure~\ref{fig:distfits}.
We choose the least-squares approach to parameter estimation for simplicity, though one might develop maximum-likelihood estimators.  We briefly discuss potential bias in our estimators below.

The residuals of the fit are used in a $\chi^2$ test to determine if the hypothetical model distribution function should be rejected.  The number of degrees of freedom (nominally the number of histogram bins) is reduced to account for the estimation of distribution parameters from the data.  We can calculate the $p$-value from the $\chi^2_n$ distribution and reject the hypothesized distribution if this measure falls below the significance level $\alpha$ (taken to be 5\%).

In contrast to a generic $\chi^2$ test, we can also use a hypothesis test based on the empirical cumulative distribution function like the Kolmogorov-Smirnov (K-S) test.  While both tests determine a $p$-value for rejecting hypotheses, we expect the K-S test to be make fewer Type II errors (failing to reject false null hypotheses) because it avoids binning data.  Disadvantages to the K-S test include a relative insensitivity to distribution tails and a much larger computation requirement: as we are using estimated parameters in the hypothesized distribution, it is necessary to determine the K-S test statistic's distribution function to find the $p$-value.  This can be done with a bootstrap Monte Carlo method, where the parameters $\hat{I_c}$ and $\hat{I_s}$ (estimated from the observations) are used to simulate a random variable following the model distribution.  For each trial $i$ in the Monte Carlo simulation, distribution parameters are estimated from the simulated data (via the preceding least-squares fit procedure) and the K-S test statistic $D_i$ is calculated.  After the completion of 1,000 trials, the distribution of $D_i$ is used to estimate the $p$-value.  We restrict the use of this form of the K-S test to example pixel locations ($a$)-($c$) because the requisite simulations cannot be calculated in a timely fashion given our computational resources.  However, the $\chi^2$ test is applicable to the entire field.

The results of the hypothesis tests for the example locations are shown in Table~\ref{tab_chisq}.  We see that neither the $\chi^2$ nor the K-S tests reject the hypothetical MRG model for these data.  The table presents similar test results for the MR model, as well as a Gaussian (G) with free parameters $\mu$ and $\sigma$.  We did not test a separate Poisson model because the data are not in the photon counting regime.  As read noise significantly affects the observed distribution of location ($a$), the model distributions consistent with Gaussian read noise provide the best fits.   However, the pure Gaussian model G is only marginally significant for locations ($a$) and ($c$), while the asymmetric distribution at ($b$) ensures rejection.  The MRG model is not rejected by either test at all three locations.
\begin{deluxetable}{llcrr@{ $\pm$ }l}
\tablecaption{Hypothesis Test Results\label{tab_chisq}}
\tablehead{\colhead{}&\colhead{}&\colhead{}&\multicolumn{3}{c}{$p$-value}\\ 
\colhead{Location} &
\colhead{Model} &
\colhead{$\chi^2/n_\mathrm{dof}$} &
\colhead{$\chi^2$ test} &
\multicolumn{2}{c}{K-S test}}
\startdata
(a) & G   & 1.61 & 0.051  & 0.26  & 0.03  \\
(a) & MR  & 2.10 & \textit{0.0087} & \textit{0.001} & 0.001 \\
(a) & MRG & 0.90 & 0.46   & 0.11  & 0.02  \\
(b) & G   & 2.51 & \textit{0.0011} & \textit{0.015} & 0.007 \\
(b) & MR  & 0.51 & 0.87   & \textit{0.001} & 0.001 \\
(b) & MRG & 0.55 & 0.83   & 0.10  & 0.02  \\
(c) & G   & 1.44 & 0.096  & 0.17  & 0.02  \\
(c) & MR  & 0.85 & 0.51   & 0.52  & 0.03  \\
(c) & MRG & 0.85 & 0.51   & 0.54  & 0.03  \\
\enddata
\tablecomments{Results of $\chi^2$ and Kolmogorov-Smirnov hypothesis tests for fits to intensity data.  G, MR, and MRG refer to the Gaussian, modified Rician, and modified Rician plus a (zero-mean, $\sigma=\sigma_\mathrm{rn}$) Gaussian models for intensity distribution.  Instances of distribution rejection are indicated in italics, and occur when the $p$-value falls below the significance level $\alpha=.05$.  The mean and variance of the G model and parameters $I_c$, $I_s$ of the MR and MRG models are estimated from the data, and $n_\mathrm{dof}$ is the number of histogram bins minus the number of estimated parameters.  The quoted uncertainty in the K-S test $p$-values corresponds to 95\% confidence.}
\end{deluxetable}

When examining the results of the $\chi^2$ hypothesis test over the image, we find that the MRG distribution hypothesis is not rejected for 90\% of the pixels.  This is slightly lower than the expected rate of $(1-\alpha)=95\%$, which is indicative of marginally significant processes which alter the intensity distribution.

Several processes may bias the results of these tests.  Read noise acts to broaden the density function, which limits the accuracy and precision of the $\hat{I_c}$ and $\hat{I_s}$ estimates in regions of low signal.  Additionally, we have chosen to neglect photon noise, as generally $\sigma_I^2\gg\langle I\rangle$.
Finite pixel size also plays a role in altering the distribution from the ideal case, as the pixels' area is a significant fraction of a speckle.  The spatially integrated speckle distribution has also been studied by~\citet{goodman75,goodman_statopt}, though we do not consider that case here. 
Another caveat arises from the non-stationarity of the process governing the wavefront distortion term $a(\vu,t)$ in the compensated imaging system, as this term is driven by the atmospheric turbulence~\citep[and references therein]{roddier81}.  These fluctuations can bias the estimates $\hat{I_c}$ and $\hat{I_c}$ and therefore reduce goodness-of-fit.
Finally, we expect second-order correlations in the sampled intensities arising from \tc\ underestimates to cause Type I errors, an increase in the test rejection rate from $\alpha$ when the hypothesis is true.

\subsection{Coherent PSF Extraction}
\begin{figure}
\plotone{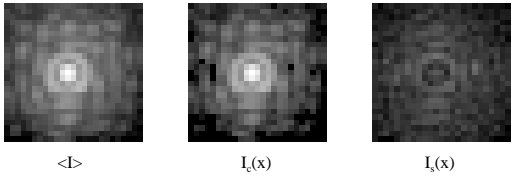}
\caption{Decomposition of the long-exposure PSF $\sim\langle I(\vx)\rangle$, reprised from Figure~\ref{fig:timeseries}, into $\hat{I_c}(\vx)$ and $\hat{I_s}(\vx)$ as in equation~(\ref{eq:mean_int}).  By definition, the quantity $I_c(\vx)$ is proportional to the static PSF.  Images are displayed with the same logarithmic scale.}
\label{fig:maps}
\end{figure}

The parameter estimation procedure above provides the spatial distributions of $\hat{I_c}(\vx)$ and $\hat{I_s}(\vx)$, the sum of which forms the long-exposure image via equation~(\ref{eq:mean_int}).  Figure~\ref{fig:maps} shows the $\hat{I_c}(\vx)$ and $\hat{I_s}(\vx)$ extracted from the observations.  As we obtain uncertainties of the estimates for these quantities via least-squares fitting, it is possible to make a crude calculation for the S/N in the estimated parameters.  Neglecting systematic biases, the median S/N over all pixels are 7.6 and 4.8 for $\hat{I_c}(\vx)$ and $\hat{I_s}(\vx)$, respectively, in data gathered in 1 minute of observing time.  It may be possible, through maximum-likelihood analysis, to develop improved estimators which avoid binning data before parameter estimation.

While $I_c(\vx)$ is a difficult quantity to measure, as it is proportional to the PSF of a system free of atmospheric turbulence, its knowledge can be valuable.  It may have utility in calibrating methods which reconstruct the long-exposure PSF ($T\gg\tc$).  With the technique of~\citet{veran97a}, data from the AO system are used to calculate a blurring kernel.  This kernel is convolved with a static PSF ($\propto I_c$) to estimate the on-axis long-exposure PSF.  This static PSF contains non-common path aberrations, and in standard practice is calibrated by an artificial point source or observations of a reference star~\citep{veran97b}.  A drawback in the artificial source case is that aberrations with high spatial frequency arising from the primary mirror are not sensed.  When using a reference star, calibrating the static PSF requires an accurate construction of the blurring kernel during the reference observation.  Obtaining $\hat{I_c}(\vx)$ by estimating the parameters of the speckle intensity distribution function (as in~\S\ref{subsec:hyp_test}) can be used to calibrate the PSF reconstruction in a manner independent of a blurring kernel estimate, though errors such as those arising from temporal non-stationarity may limit the accuracy.  Finally, we note that the reference star calibration method presented in~\citet{veran97b} and the speckle distribution function method above require that the calibration target be a point source.  If incoherent light is added as from a companion or extended emission, the speckle distribution will be altered from the MR.  An analytical expression for the intensity distribution function in this case is given by~\citetalias{aime&soummer04}.

\section{CONCLUSIONS}\label{sec:conclusions}

We have shown observations of speckle intensity over timescales of $\sim$1 min are consistent with modified Rician statistics through $\chi^2$ and Kolmogorov-Smirnov goodness-of-fit tests.  In high-contrast imaging for planet detection, the variance of the integrated intensity at the location of the planet sets the noise level against which the signal must be detected.  Central to understanding this noise level are the variance of intensity $\sigma_I^2$ and the speckle decorrelation time \tc.
\citetalias{aime&soummer04} highlight the efficacy of coronagraphs in reducing $\sigma_I^2$ by reducing terms containing $I_c$.  For example, suppose $I_c=I_s=1$.  From equation~(\ref{eq:modric_var}), $\sigma_I^2=3$.  With a coronagraph, $I_c=0$ and $\sigma_I^2=1$, a reduction by a factor of 3 in variance.  However, a complete understanding of speckle fluctuations in coronagraphic imaging experiments will require characterization of \tc, as well as understanding the effects of non-stationarity in $I_c$ and $I_s$ due to changes in the atmosphere and quasi-static wavefront errors in the optical system.  Analyses of observed intensity distributions over timescales longer than 1 min are left to future work.

We note that the dark speckle imaging technique relies on analysis of the observed intensity distribution at each location \vx~\citep{labeyrie95,boccaletti98a}, and has traditionally assumed exponential statistics for intensity.  Here we have demonstrated that probability distributions related to the modified Rician are applicable in the case of moderate wavefront correction and thus may be incorporated into this technique.  \citetalias{aime&soummer04} give speckle distributions for additive planet light and photon-counting conditions which can be used to this end.

Additionally, we have presented a technique where $I_c(\vx)$ and $I_s(\vx)$ are estimated from the observed distributions of intensity.  Knowledge of the unaberrated PSF, proportional to $I_c(\vx)$, can be applied to the calibration of PSF reconstruction algorithms~\citep{veran97a}.  This application may spur the further development of estimators for $I_c$ and $I_s$ given an observed speckle intensity distribution.

\acknowledgements
This work was supported in part by the NSF Science and Technology Center for Adaptive Optics, managed by the University of California at Santa Cruz under cooperative agreement AST-9876783.
The authors would like to thank the staff of UCO/Lick Observatory, as well as Don Gavel, Paul Kalas, James Lloyd, Christian Marois, Dave Palmer, Marshall Perrin, and Peter Tuthill for their input and assistance.
Special thanks are due to R\'emi Soummer for providing advanced copies of his elegant papers on the statistical properties of speckles and for invaluable suggestions and advice.
This publication makes use of data products from the Two Micron All Sky Survey, which is a joint project of the University of Massachusetts and the Infrared Processing and Analysis Center/California Institute of Technology, funded by the National Aeronautics and Space Administration and the National Science Foundation.

\bibliography{}
\end{document}